\documentclass[twocolumn,prl]{revtex4}
\usepackage{amsmath}
\usepackage{amssymb}
\usepackage{latexsym}
\usepackage{graphicx}
\usepackage{gensymb}
\usepackage{bm}

\def\figno#1{Fig.~\ref{fig:#1}}

\def\kb{k_{\scriptscriptstyle\rm B}}

\def\figW{80mm}
\begin{document}
\title{
Fluctuation spectroscopy of surface melting of ice without, and with impurities
}
\author{Takahisa   Mitsui\footnote{E--mail:~{\tt      mitsui@phys-h.keio.ac.jp}.} and 
Kenichiro Aoki\footnote{E--mail:~{\tt ken@phys-h.keio.ac.jp}.} }
\affiliation{Research and Education Center for Natural Sciences and
  Dept. of Physics, Hiyoshi, Keio University, Yokohama 223--8521,
  Japan}
 \begin{abstract} 
   Water is ubiquitous, and surface properties of ice has been studied
   for some time, due to its importance.  Liquid-like layer (LLL) is
   known to exist on ice, below the melting point. We use surface
   thermal fluctuation spectroscopy to study LLL, including its
   thickness, for pure ice, and for ice with impurities. We find that
   the properties of LLL are experimentally those of liquid water,
   with thickness much smaller than previous results. We also find
   that impurities cause LLL to be thicker, and be quite
   inhomogeneous, with properties depending on the dopant.
 \end{abstract}
\maketitle 
\par

The importance of understanding the surface properties of ice in air
can not be understated, which is crucial for the clarification of the
melting, freezing process itself, as well being important to a broad
range fields in science \cite{Jellinek,DRW2006,Bartels}.
Theoretically, the existence of LLL (also called
``quasi-liquid layer'', or QLL) has been discussed since 19th century,
and its properties, including the existence, has been studied from
various points of view --- thermodynamic
arguments\cite{Weyl,Fletcher,ES}, and more recently, state of the art
molecular dynamics simulations have been
employed\cite{Kroes,Nada2000,MD1,Ikeda2004,Pfalzgraff2011,Shepherd2012,MD2,Mohandesi2018,Pickering2018}. Due
to its importance, and its experimental difficulty, surface melting of
ice has been analyzed experimentally using a multitude of methods,
such as ellipsometry\cite{BN,FYK,Furukawa1997}, X-ray
diffraction\cite{X1}, proton backscattering\cite{pBackscattering},
photoelectron spectroscopy\cite{PES}, atomic force
microscopy\cite{AFM1,AFM2,AFM3}, and surface sum-frequency generation
(SFG) spectroscopy\cite{Sanchez2017,Smit2017,IRS}.  For a number of
reasons, including the thinness of LLL, which can be nm order or
smaller, the study of LLL remains to be an experimentally challenging
problem. This difficulty is evidenced by the thickness measurements of
LLL of ice, which vary by orders of
magnitude\cite{ELD,Furukawa1997,Bartels}.

In this work, we optically measure the surface thermal fluctuations of
LLL on ice, in air. By measuring the thermal motions of the molecules
directly, in addition to clearly differentiating the liquid and solid
phases, the properties of the material become
apparent\cite{Levich,Bouchiat,Jackle,Langevin,Cicuta2004,Sagis}.
Such a method, while not previously applied to surface melting, has
proven to be effective in understanding the properties of various
liquids, complex fluids, and viscoelastic
materials\cite{MitsuiOnion,Tay,MA1,Pottier11,Pottier2015,AMPTEP}.  This
experimental method for measuring the properties of LLL, distinct from
previous methods, enables us to obtain a different perspective on LLL.
The surface thermal fluctuations, which are spontaneous, reveal the
properties of the material underneath, providing information whether
LLL has the properties of water in the bulk, in addition to the
behavior of its thickness. 
The importance of thermal fluctuations of the LLL were recognized, and
their properties were recently analyzed using
simulations\cite{Benet2016}.  There, it was found that the surface
fluctuations are similar to those of water--vapor interfaces, with the
fluctuation spectra strongly affected by the thickness of LLL,
consistent with our results.
Also, it has been found using SFG
spectroscopy\cite{Sanchez2017,Smit2017,IRS}, that the surface of LLL
behaves similarly to bulk supercooled water, seemingly consistent with
our results.
We find that LLL is much thinner than most of the previous
experimental results, and that the additions of impurities at ppm
levels  thicken LLL.  Furthermore, by scanning the surface at
$\mu$m level resolution, we directly observe that impurities cause
inhomogeneities, with their properties depending on the dopant.

\begin{figure}[htbp]
  \centering
   \includegraphics[width=\figW,clip=true]{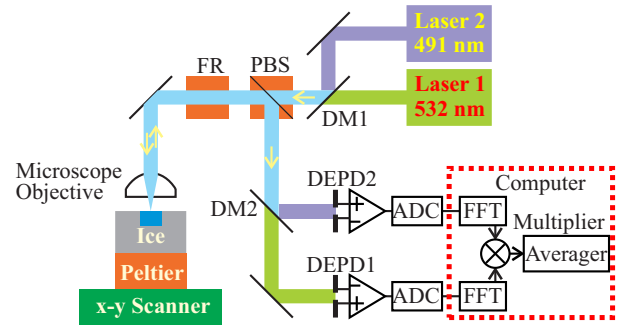}
   \caption{{ Experimental setup:} Linearly polarized laser light with
     wavelengths 532, 491\,nm are shone on the sample ice surface, 
     with powers 340, 270$\,\mu$W at the surface. 
     The reflected light is directed to two dual-element photodiodes
     (DEPD1,2) corresponding to Laser 1,2. 
     The differences in the light beam powers in the
     DEPDs are digitized using analog-to-digital converters (ADC), 
     Fourier transformed (FFT), and the averaged
     correlation is computed (averager). 
     Faraday rotator (FR) is used
     to rotate the polarization of the light by $\pi/4$, each way, so
     that the polarized beam splitter (PBS) reflects the light from
     the sample.  Dichroic mirrors (DM1,2) are used to merge, and
     separate the light with different wavelengths. 
   }
  \label{fig:setup}
\end{figure}
The experimental setup is shown in \figno{setup}: Light is shone on
the surface of ice, with LLL expected to be on it. The reflected light
is detected by the dual-element photodiode (DEPD).
The surface acts as a partial mirror, and the two elements in DEPD
produce the same photocurrent, if the surface is not fluctuating. The
surface fluctuates thermally, and produces fluctuations in the
photocurrent difference, whose power spectrum is the inclination fluctuation
spectrum of the surface, up to a constant. The measurement is
performed using two light sources with different wavelengths, in order
to use correlation analysis to statistically reduce the extraneous
noise\cite{MA1}, to orders of magnitude below the shot-noise level,
often referred to as the ``Standard Quantum Limit''\cite{quantumOptics}.
The beam radius (waist) at the sample, $w$,  is $1.2\,\mu$m, and the sample can be
moved horizontally in two dimensions by $15\,\mu$m in each direction,
allowing us to scan the surface.
Recently, LCM-DIM (laser confocal microscopy combined with
differential contrast microscopy) has been developed to optically
observe ice surfaces\cite{Asakawa2016,Murata}. Our method measures the
thickness of LLL, which is not measurable in LCM-DIM. LCM-DIM has been
used to measure transient properties of surface structures, in
contrast to our thermal equilibrium measurements, and while both
methods are optical measurements, we find these methods complementary.
Due to the breadth of interest, importance of the surface melting
phenomena, and the difficulties in measuring their properties, various
experimental approaches have been brought to bear on this problem.  We
believe that it is important to investigate the phenomena from
different perspectives, both experimental and theoretical, to clarify
the inner workings underlying them.  One important character of our
approach is that it makes optical measurements of only the spontaneous
surface fluctuations, with no external stimulation, and is hence
minimally invasive. The time-scales for our measurements at relatively
longer time scales, at least order of tens of seconds. The approach is
suited for measurements of thermal equilibrium properties, and not for
observing transient effects. The measurements made in this work are
stable on the order of hours.

The liquid sample to be frozen was put into the stainless steel
container (diameter 3.5\,mm, depth 1.5\,mm), and thermoelectrically
cooled. Due to supercooling, the liquid froze at around $-20$\degree
C, without, and with impurities.  For pure water, and NaCl solution
cases, the temperature of the liquid was then raised to 0.07\degree C,
and kept at that temperature till the water layer on ice was about
0.5\,mm, which took around 30 minutes. Next, the sample liquid
temperature was lowered gradually to $-0.03$\degree C, taking over an
hour.
After this, the measurements were then taken, at various temperatures.  
For the Volvic sample, the same procedure led to too much
precipitation of minerals, leading to unwanted noise in the
measurement due to the unevenness of the surface.  Therefore, to
reduce the precipitation, the temperature of the sample was raised to
to $-2.4$\degree C after freezing the supercooled Volvic water,  and kept
constant for over an hour, after which measurements at various
temperatures were taken.  
Water sample was purified using the EMD
Millipore 
Purification System (Merck Millipore, Germany).

\begin{figure}[htbp]
  \centering
   \includegraphics[width=\figW,clip=true]{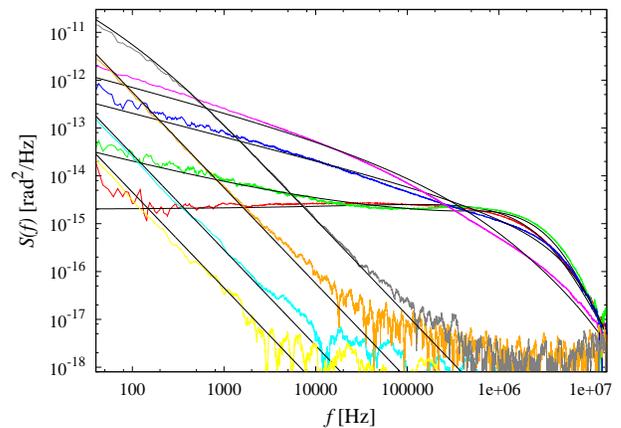}
   \caption{{ Surface thermal fluctuation spectra of water layer of
       various thicknesses:}  Fluctuation spectra for supercooled
     water ($\Delta T=17.89\,$K, red), and LLL with
     $(\Delta T\,[{\rm K}],h\,[{\rm m}])=$ $(0.003,3.0\times10^{-6})$
     (green), $(0.004,6.0\times10^{-7})$ (blue),
     $(0.006,2.5\times10^{-7})$ (magenta), 
     $(0.014,1.0\times10^{-8})$     (grey), 
     $(0.024,3.6\times10^{-9})$ (orange),
     $(1.39,1.4\times10^{-9})$ (cyan), 
     $(24.89,1.0\times10^{-9})$
     (yellow). 
     Corresponding theoretical spectra for water with finite depth are
     also shown(black), and agree well with the measured spectra.}
  \label{fig:spectra}
\end{figure}
Our experimental setup measures the thermal fluctuation spectra of the
averaged inclination within the beam spot on the sample\cite{MA1},
\begin{equation}
  \label{eq:spectrum}
  S(f)=\int_0^\infty dk\,k^3e^{-w^{2}k^2/4}F(k,f,h)\quad.  
\end{equation}
Here, $f$ is the frequency, and $F(k,f,h)$ is the spectral function of
the thermal fluctuations of the fluid surface, and depends on
LLL thickness, $h$\cite{Jackle}.
The spectrum is uniquely determined by the bulk properties of water
(density, surface tension, viscosity), beam size, $h$, and the sample
temperature. Gravitational effects are negligible at our sample size.
Some examples of spectra for LLL, and water at various
$h,\Delta T=T_{\rm m}-T$ are shown in \figno{spectra}.  $T$ is the
temperature of the sample surface, and $T_{\rm m}$ is the bulk melting
temperature of ice. It can be seen that the spectral shape depends
strongly on $h$, and the experimental spectra agree quite well with
the theoretical spectra of surface thermal fluctuations of water with
finite thickness.  The known bulk properties of supercooled
water\cite{water1,water2,water3,CRC} were used to compute the
theoretical spectra.  We note that surface thermal fluctuations of
solid ice without LLL should not only be much smaller, but behave as
$1/f$\cite{Tay,MA1}, which is incompatible with the measurements. For
LLL with $h\gtrsim10\,$nm, the spectral shape and the magnitude are
sensitive to the fluid properties of LLL, density, surface tension,
and viscosity, as well as its thickness, in our experimental setting.
In particular, in all the spectra analyzed, the viscosity inferred
from the spectrum is consistent with the known physical properties of
supercooled water.  For $h\lesssim10\,$nm,
the theoretical spectrum behaves as
\begin{equation}
  \label{eq:shallow}
  S(f)\simeq{16\over3\pi^3}{\kb T\over\eta w^6}{h^3\over f^2}\quad, 
\end{equation}
where $\eta$ is the viscosity of (supercooled) water, $\kb$ is the
Boltzmann constant, and $T$ is the temperature.  This $1/f^2$ behavior
is indeed observed for thinner LLL in \figno{spectra}.  While the
spectral magnitude is quite sensitive to the thickness, the spectral
shape is rather insensitive to the properties of LLL.

\begin{figure}[htbp]
  \centering
  \includegraphics[width=\figW,clip=true]{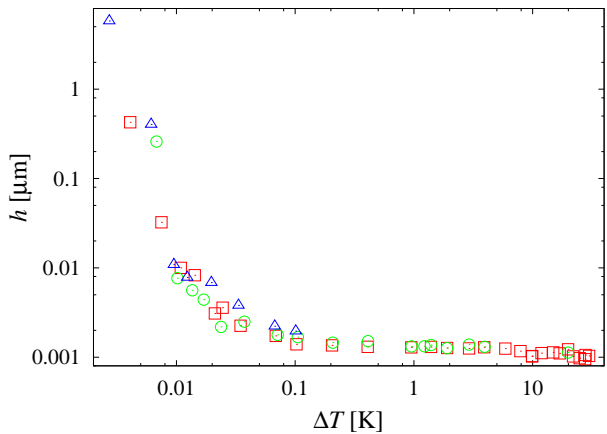}    
  \caption{{ The temperature dependence of the thickness of LLL for
      pure ice:} The three types of points ($\bigcirc,\triangle,\Box$)
    correspond to three data sets taken with three different samples,
    on three different days, which are seen to be consistent within
    error. }
  \label{fig:thickness}
\end{figure}
The dependence of $h$ for LLL of pure ice on the temperature is shown
in \figno{thickness}.  $h$ was estimated from the surface thermal
fluctuation spectra (\figno{spectra}): 
For $h\gtrsim10\,$nm, theoretical spectral shape was fit to
the measurements, and for smaller $h$, the measured spectrum was
normalized using the shot-noise level.  The relation between $h$ and
$\Delta T$ has been studied previously by a number of authors, using
various other methods, and $h$ differs by orders of magnitude,
depending on the method used\cite{ELD,Furukawa1997,Bartels}.
Our results extend over a much wider range of temperatures than that
previously covered with any one method. 
The results in \figno{thickness} are consistent between measurements
of different samples taken on different days, and  also 
between measurements taken with the temperatures in  ascending, and
 descending sequences.  This strongly suggests that the results are
equilibrium properties.
Our results for $h$ are smaller
than those previously measured for $\Delta T<1$\,K, close to the
bulk melting temperature. Previous results are almost non-existent for
$\Delta T>10\,$K, except for photoelectron emission spectroscopy study
of LLL of ice in pure water vapor\cite{PES}. Compared with this, our
results for $h$ are of the same order but slightly larger in
this temperature range.
On the theoretical side, some thermodynamic considerations predict
much thicker\cite{Fletcher}, and also thinner\cite{ES} LLL.
Recent molecular dynamics simulation results exist for
$\Delta T\gtrsim1\,$K\cite{MD1,MD2,Shepherd2012,Pickering2018,Benet2016}, with thicknesses smaller than
ours, but consistent within a factor of few.
It should be noted that  previous literature with thinner LLL, both
theoretical and experimental\cite{PES,ES,MD1,MD2}, deal with ice in
pure water vapor, without air, unlike our results, which might be the
reason for the difference.
For thin layers, viscosity might be larger than its
value in bulk, though whether it is, and by how much if so, is an
unsettled question\cite{visc1,visc2,visc3,DRW2006}.  Since the surface
thermal fluctuation spectrum behaves as $\sim h^3/(\eta f^2)$ for thin
LLL, larger viscosities lead to larger $h$ values in our results
(\figno{thickness}), so that a dramatic rise in the viscosity for smaller $h$ seems
unlikely.

\begin{figure}[htbp]
  \centering
  \includegraphics[width=\figW,clip=true]{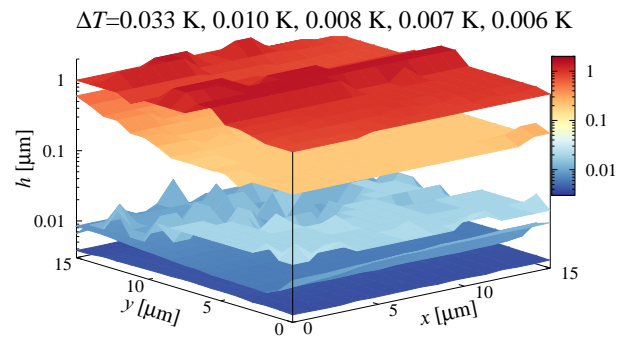}    
  \caption{{ Spatial distribution of LLL thickness of pure ice:}
    $h$, at four temperatures,
    $\Delta T= 0.033, 0.010, 0.008, 0.007, 0.006\,$K.  $h$ is larger
    at lower temperatures (larger $\Delta T$).  $h$ distributions are
    seen to be relatively uniform, with slight variations. }
  \label{fig:scanWater}
\end{figure}
An important questions is whether LLL is homogeneous: Our measurement
system allows for scanning at $\mu m$ level, since the light beam is
focused, with $w=1.2\,\mu$m. In \figno{scanWater}, the dependence of the
$h$ on the surface location is shown, for the surface of pure ice, at
different temperatures.
$h$ values are seen to be reasonably uniform in the scanned region,
for LLL of pure ice.  For small $\Delta T$, $h$ is strongly dependent on
the temperature (\figno{thickness}), so some variation is visible.
Molecular dynamics simulations exist that suggest  the  transition from
liquid to solid is not sharp for
LLL\cite{Pickering2018,Benet2016}. The
properties we found through the surface thermal fluctuations are
consistent with a thin layer having the properties of bulk water, as
seen in \figno{spectra}. We expect this not to preclude a gradual
transition from the liquid-like to the solid-like structure at the
bottom, as long as waves with amplitudes larger than those at the
liquid-gas interface do not propagate in this transition layer. It
would be interesting to investigate if more precise measurements, and
further information can be extracted from surface thermal fluctuation
measurements, especially when LLL is thin, around $1\,$nm or less.
Thermodynamic arguments combined with microscopic simulation suggest,
interestingly, that the ice-liquid interface has a roughness, at
atomic scales\cite{Qiu2018}. However, since our measurements are
averaged within the beam at the $\mu$m order, such roughness is not
observable.

\begin{figure}[htbp]
  \centering
\def\figW{\textwidth}
\begin{minipage}{0.49\linewidth}
\includegraphics[width=\figW,clip=true]{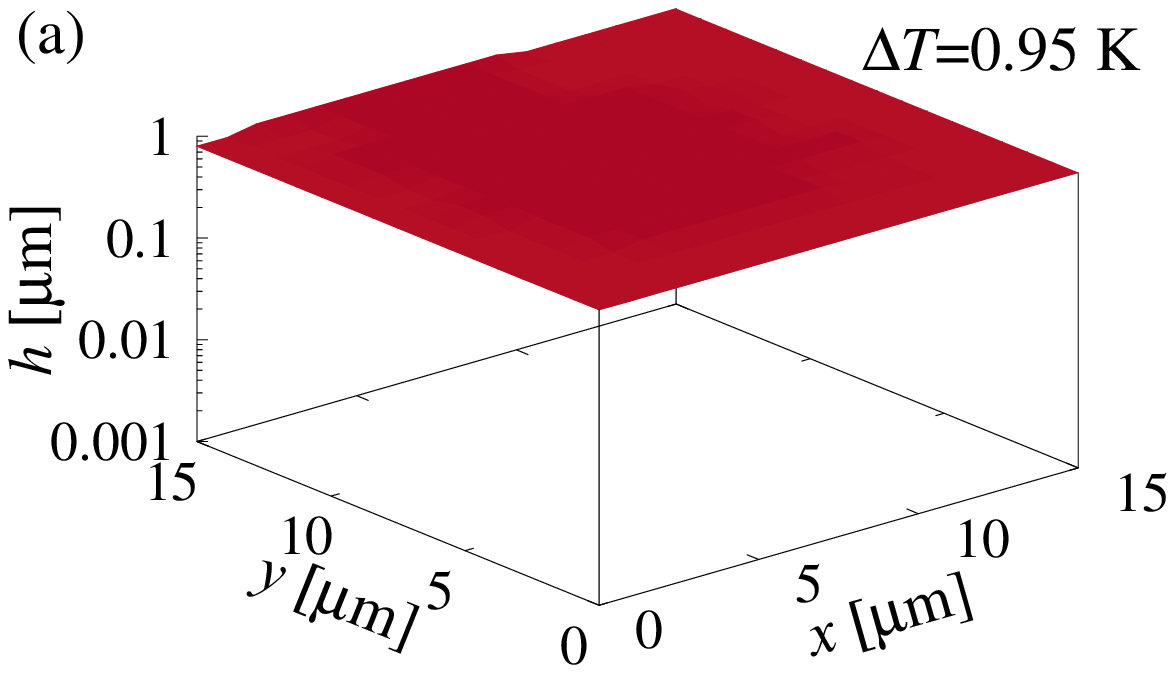}
\includegraphics[width=\figW,clip=true]{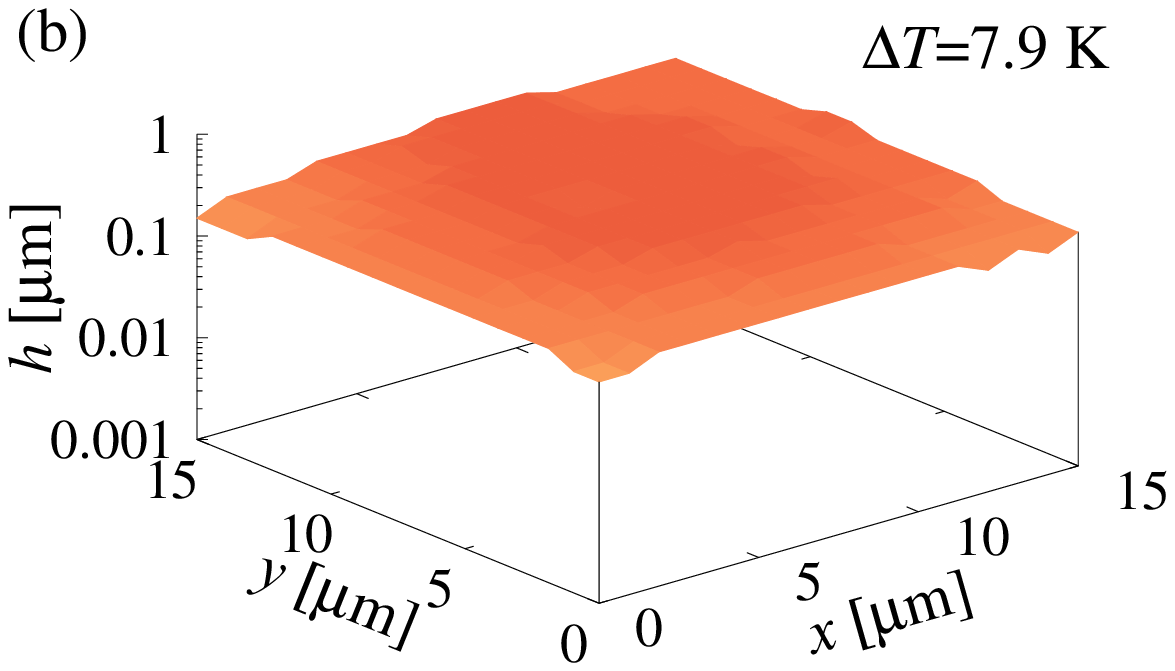}
\includegraphics[width=\figW,clip=true]{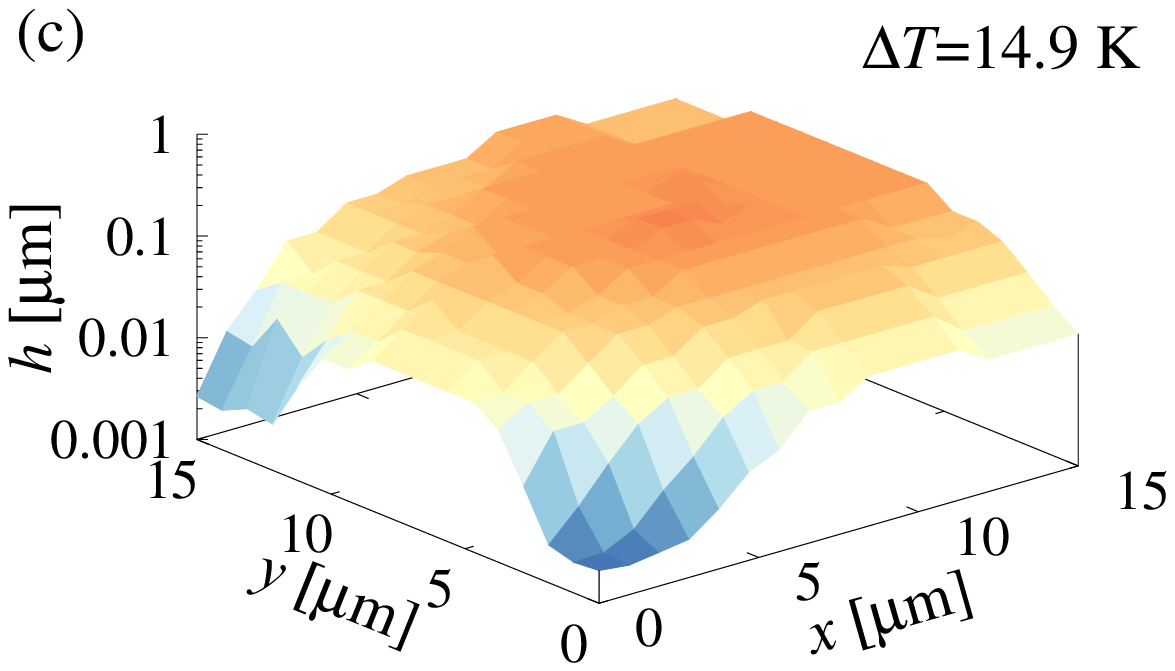}
\includegraphics[width=\figW,clip=true]{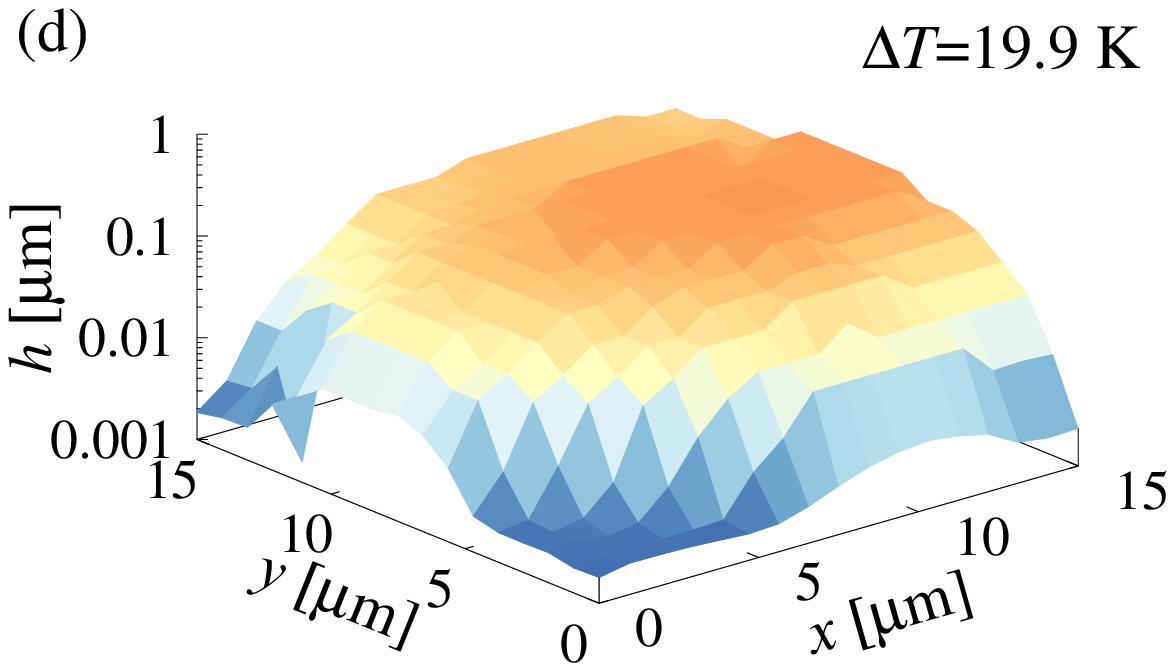}
\includegraphics[width=\figW,clip=true]{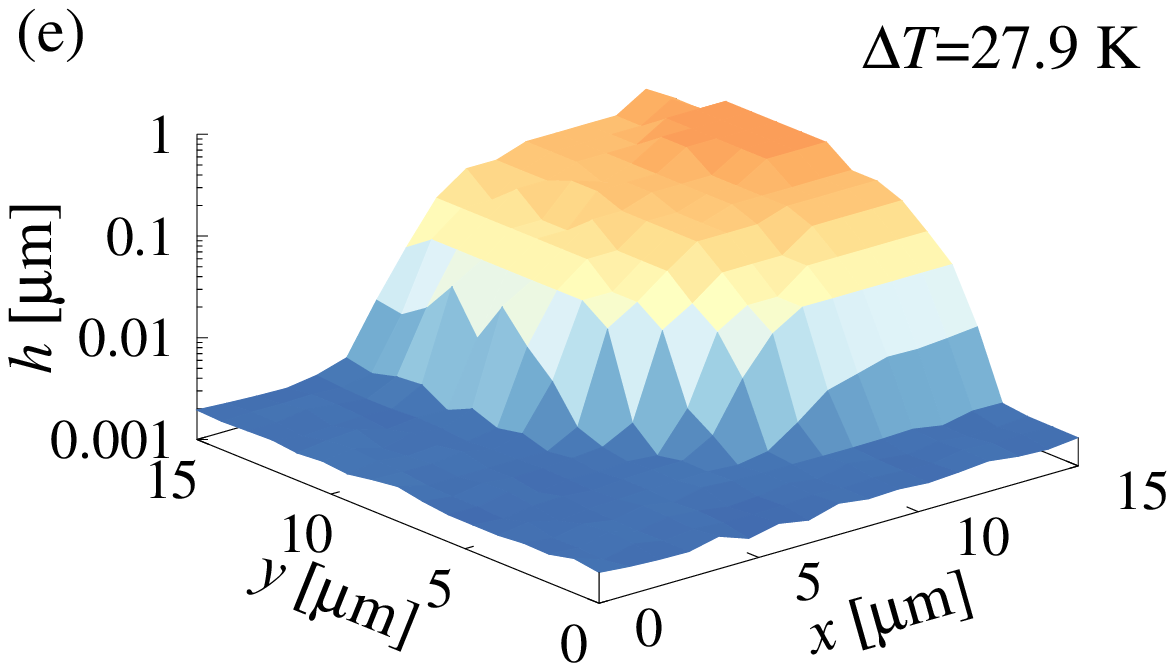}
\end{minipage}
\begin{minipage}{0.49\linewidth}
\includegraphics[width=\figW,clip=true]{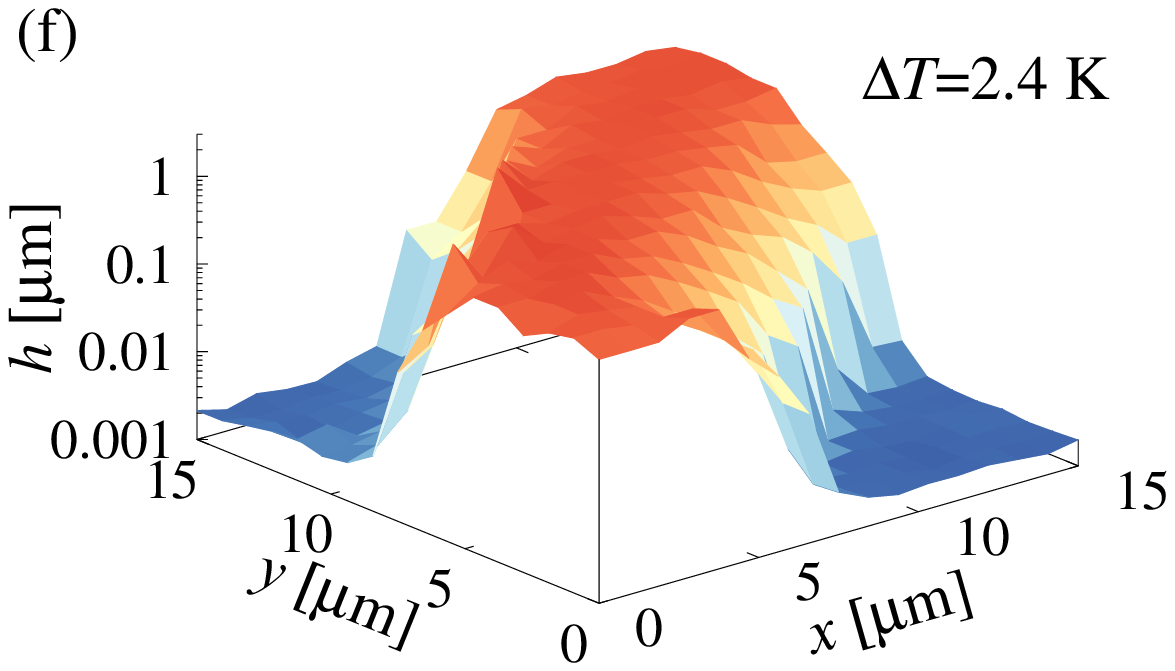}
\includegraphics[width=\figW,clip=true]{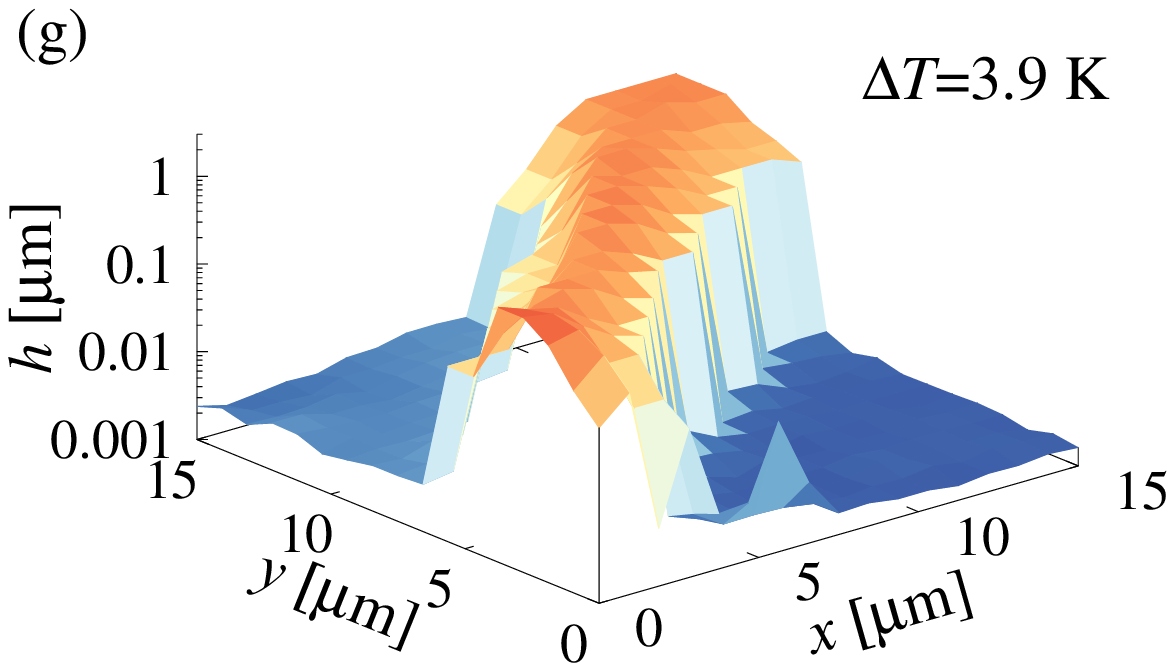}
\includegraphics[width=\figW,clip=true]{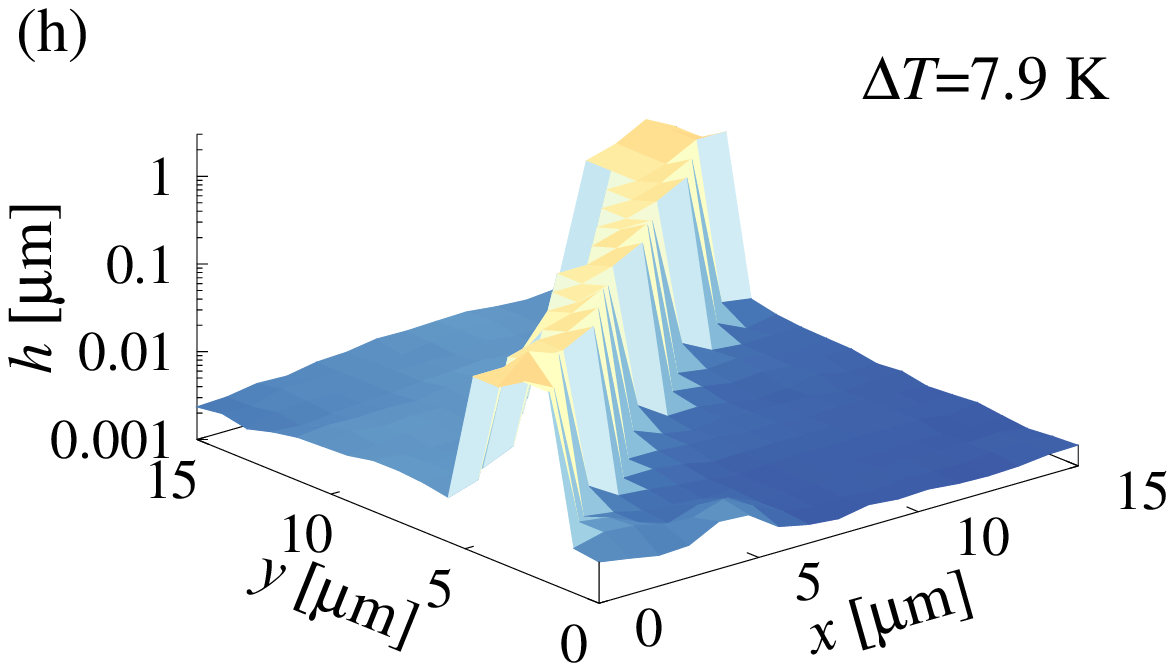}
\includegraphics[width=\figW,clip=true]{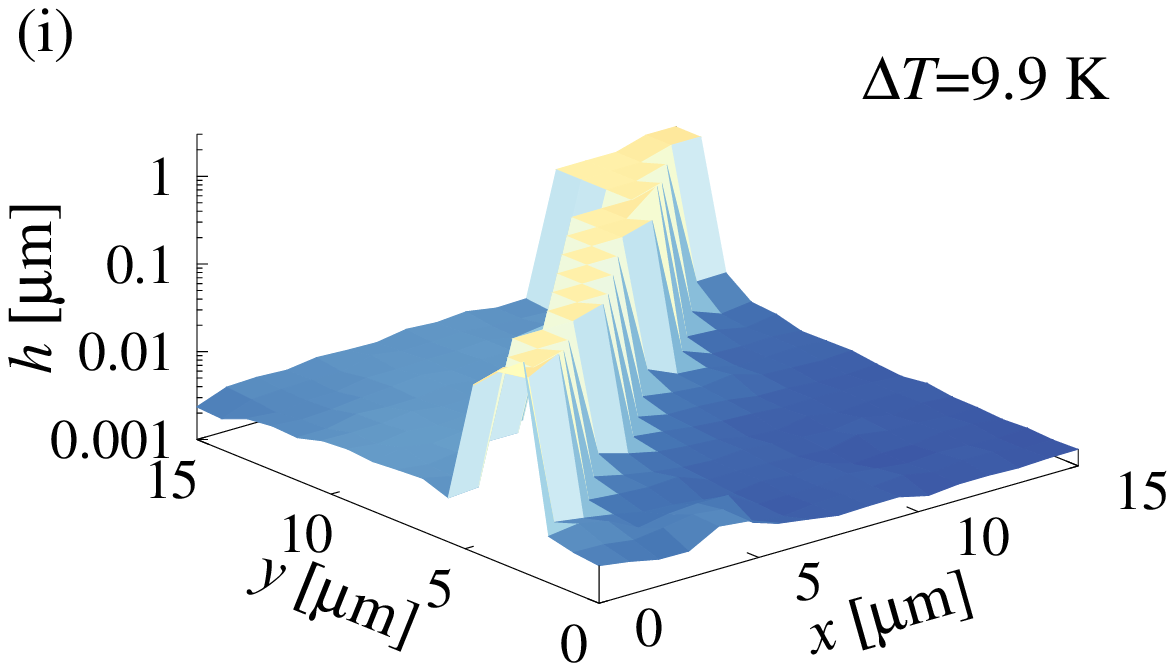}
\includegraphics[width=\figW,clip=true]{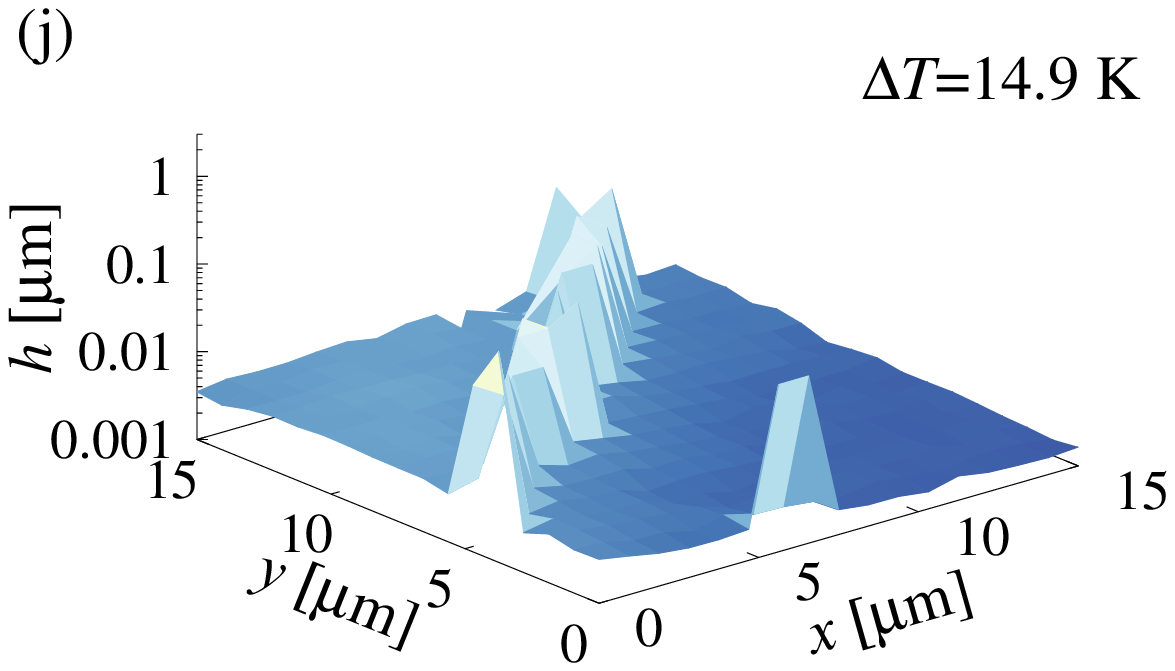}
\end{minipage}
\caption{
Spatial distribution of
  $h$ for ice with NaCl ((a)---(e)), and Volvic ((f)---(j)),
  as the temperature is lowered, for the same area. 
}
  \label{fig:scanImp}
\end{figure}
Theoretically, impurities can greatly affect the overall thickness of
LLL\cite{imp1,imp2}, and are perhaps the major cause of their large observed
disparities\cite{DRW2006}. To study the effect of impurities, in
\figno{scanImp}, the spatial dependences of $h$ are shown for frozen
NaCl solution (10\,ppm by weight before freezing), and Volvic (water with
minerals roughly 60\,ppm by weight)\cite{Volvic} at various
temperatures, as they are being cooled. 
Such direct observations of the inhomogeneities have been observed for
the first time.
 We chose Volvic, which
contains various minerals, as a model of water in a natural
setting.
There are clear qualitative differences from the properties of LLL of
pure ice, and also between LLL with different impurities. First, we
see that in both cases, LLL with $h>0.1\,\mu$m exists for
$\Delta T>1\,$K, in contrast to that of pure ice, seen in
\figno{thickness}.  Furthermore, unlike pure ice LLL, $h$
distributions are quite inhomogeneous. There is also a distinct
difference between the effect of two impurities: A reasonably thick
LLL exists for frozen NaCl solution down to temperatures much lower
than that for Volvic, which is not as inhomogeneous as the latter. For
Volvic, channels of LLL form, which have the bulk properties of water,
that grow narrower and shallower as the temperature
lowers. Interestingly, vein like structures have been observed in
glaciers\cite{RH,DRW2006}.
The cause for the distinct difference between ice with NaCl and Volvic
can perhaps be attributed to the difference in the solubility of the
impurities.  Minerals within Volvic are not as soluble in water as
NaCl, in general, probably leading to more inhomogeneities and precipitation.
We have scanned the surface of ice with glucose, and have found that
the behavior is similar to that of NaCl solution.
Results for ice with NaCl at various concentrations, and frozen Evian
water\cite{Evian}, and are also consistent with the above
considerations.
The concentration per unit area of the NaCl solution in
\figno{scanImp} can be estimated to be $90\,\mu\cdot$mol/m$^2$.  The
observed thicknesses of LLL seem roughly consistent with those of the
theory\cite{imp1}, though at lower temperatures, the inhomogeneity of LLL
should be taken into account.

An interesting classic problem is the ability to skate on ice, or the
slipperiness of ice. %
The most classic explanation for its cause is pressure
melting. However the theory is insufficient in explaining the
ability to skate, except very close to the melting point, due to the
pressures required\cite{DRW2006}.
 One explanation is the existence of a relatively
thick LLL. We found that pure ice only supports LLL with thicknesses
under 1\,nm for $\Delta T>1$\,K. However, in a natural setting, water
inevitably contains a certain amount of impurities. Our results show
that impurities thicken LLL, and create inhomogeneities, which can
contribute to the slipperiness of ice. 
One observation from our results is that  frozen mineral water
supports surface melting in the form of veins, which essentially
disappear for temperatures below $-10$\degree C. This is intriguing
considering that optimal skating conditions are considered to be between $-5$ to
$-9$\degree C\cite{oly}.
Frictional heating\cite{DRW2006}, and other mechanisms have also been
recently suggested\cite{Murata,Weber}, and further investigation
remains.

\vskip0mm\noindent
{\bf Acknowledgments }
The authors were supported in part by the Grant--in--Aid for
Scientific Research (Grant No.~15K05217) from the Japan Society for
the Promotion of Science (JSPS), and a grant from Keio Gijuku Academic
Development Funds.  
\end{document}